\documentclass[twoside,12pt]{article}
\usepackage{cite}
\usepackage{color}
\usepackage{graphicx}
\newcommand{\epem}         {e^+e^-}
\newcommand{\nunubar}      {\nu\bar{\nu}}
\newcommand{\Egamma}       {E_{\gamma}}
\newcommand{\Ebeam}        {E_{\mathrm{beam}}}
\newcommand{\Etot}         {E_{\mathrm{tot}}}
\newcommand{\Etotsq}       {E_{\mathrm{tot}}^2}
\newcommand{\MeV}          {{\mathrm{MeV}}}
\newcommand{\pb}           {{\mathrm{pb}}}
\newcommand{\Mchi}         {M_{\chi}}
\newcommand{\MU}           {M_U}
\newcommand{\Mpi}          {M_\pi}
\newcommand{\Cchi}         {C_\chi}
\newcommand{\fe}           {f_e}
\newcommand{\Beeann}       {B^{ee}_{\mathrm{ann}}}

\newcommand{\bi}   {\begin{itemize}}
\newcommand{\ei}   {\end{itemize}}
\newcommand{\be}   {\begin{enumerate}}
\newcommand{\ee}   {\end{enumerate}}
\newcommand{\bcen} {\begin{center}}
\newcommand{\ecen} {\end{center}}
\newcommand{\beq}  {\begin{equation}}
\newcommand{\eeq}  {\end{equation}}

\newcommand{\etal} {{\em et al.}}

\begin{document}
\rightline{NUHEP-EX/07-01}
\begin{center}
{\Large An Experiment to Search for Light Dark Matter
\vskip 4pt
in Low-Energy $ep$ Scattering}
\vskip 10pt
Sven Heinemeyer$^{1}$,
Yonatan Kahn$^{2}$,
Michael Schmitt$^{2}$,
Mayda Velasco$^{2}$ \\
\vskip 5pt
{\sl
$^{1}$Instituto de Fisica de Cantabria (CSIC-UC), Santander, Spain\\
$^{2}$Northwestern University, Evanston, Illinois, USA
}
\vskip 10pt
December 12, 2007
\end{center}
%-------------------------------------------------
\vskip 20pt
\begin{center}
{\bf Abstract}
\vskip 5pt
\parbox[c]{0.85\textwidth}
{\small
Anomalous production of low-energy photons from the galactic
center have fueled speculations on the nature and properties
of dark matter particles.  In particular, it has been proposed
that light scalars may be responsible for the bulk of the
matter density of the universe, and that they couple to 
ordinary matter through a light spin-$1$ boson.  If this
is the case, then such particles may be produced in the
quasi-elastic low-energy scattering of electrons off protons.
We present a proposal for an experiment to search for this
process and assess its viability.
}
\end{center}
\vskip 20pt
%
%%%%%%%%%%%%%%%%%%%%%%%%%%%%%%%%%%%%%%%%%%%%%%%%%%%%%%%%%%%%%%%%%%%%%%
\section{Introduction}
\par
The INTEGRAL data provide evidence for an anomalous production
of low-energy gammas from the galactic center~\cite{INTEGRAL}.
The excess is largest in the region at and below $\Egamma = 511$~keV
leading to the conclusion that positron-electron annihilation
is responsible.  The question becomes: where are the extra
positrons coming from?  One answer is that they are produced
in the annihilation processes of dark-matter particles,
which we will write as $\chi\chi^* \rightarrow \epem$,
where $\chi$ denotes the dark-matter~(DM) particle.
\par
A phenomenological model has been developed which posits
a light dark-matter particle~$\chi$ and a new spin-1 gauge boson~$U$
which mediates interactions between $\chi$ and ordinary 
matter particles, such as electrons~\cite{BoehmFayet}.  
The couplings of the $U$-boson to standard model fermions is 
not specified by any theoretical first principles; rather, 
they are left as parameters to be constrained by the known
dark-matter abundance and processes studied in the laboratory.
At a minimum, there must be a non-zero coupling to the
dark matter particles as well as to electrons, a fact which
we exploit in our proposal to observe $U$ bosons in the
laboratory.
\par
The experiment we propose makes use of a low-energy
electron beam ($\Ebeam \approx 40~\MeV$) and a fixed hydrogen
gas-jet target.  We describe two versions: the first is
rather simple and requires a minimum of resources, and 
the second is more elaborate allowing
for a greater sensitivity.  We attempt to compare the
sensitivity and potential information gained from our
proposed experiment to other suggested avenues of research, 
such as the study of $\epem$ collision data~\cite{epem}.
\par
This paper is organized as follows.  We begin with a brief
review of the evidence for the low-energy gamma-ray excess 
from the galactic center, followed by a resume of the 
light dark matter models recently discussed in the literature.
Next we explain the process of interest in general
terms, followed by a detailed description of the two
experimental set-ups to study this process.  We make
a comparison with other proposals followed by a summary
of our work.

%%%%%%%%%%%%%%%%%%%%%%%%%%%%%%%%%%%%%%%%%%%%%%%%%%%%%%%%%%%%%%%%%%%%%%
%
\section{Gamma-Ray Excess}

In 1972, Johnson \etal~\cite{Johnson1972} used a balloon-hoisted NaI scintillation 
telescope to detect a 511~keV line emanating from the center of the galaxy. 
After accounting for 511~keV radiation resulting from cosmic rays and 
positron annihilation in the upper atmosphere, they concluded that this line 
was significantly stronger than the cosmic background radiation from other 
directions at similar energies. Leventhal \etal~\cite{Leventhal1978} revisited this 
phenomenon in 1978 and identified the source of the radiation as positron 
annihilation; specifically, they concluded that positrons in the galactic 
center form both para-positronium, which annihilates to give the 511~keV line, 
and ortho-positronium which contributes a spectrum of excess low-energy 
radiation at and below $511$~keV. Further experiments over the past 30 years, 
and most recently the INTEGRAL satellite, have refined the value of the 
observed photon flux and identified the majority of the radiation as coming 
from the galactic bulge~\cite{INTEGRAL,Cheng1997,Purcell1997,Vedrenne2003}. 
Currently, the explanation of 
annihilating positronium is well-accepted as the source of the radiation, but 
the source of the positrons themselves remains unclear. Several 
possibilities involving relatively well-known astrophysical phenomena have 
been put forth, including radioactive nuclei from supernovae~\cite{Milne2002},
gamma-ray bursts~\cite{Bertone2006}, pulsars~\cite{Sturrock1971},
black holes~\cite{Ramaty1989}, and cosmic rays~\cite{Kozlovsky1987}, but 
these models have difficulty accounting for the morphology and high intensity 
of the photon flux except under rather restrictive assumptions.

To resolve these problems, several more exotic explanations from particle 
physics have been proposed. A partial list compiled by Sizun \etal~\cite{Sizun2006}
includes Q-balls, relic particles, decaying axinos, primordial black holes, 
color superconducting dark matter, superconducting cosmic strings, dark 
energy stars, moduli decays, and annihilating light dark matter particles. 
The latter explanation, in which dark matter particles annihilate in the 
galactic bulge to form $\epem$ pairs, has been able to account for both 
the morphology and intensity of the line as long as the dark matter particles 
are light ($< 100~\MeV$)~\cite{Boehm2004a,Boehm2004b,Boehm2004c,Boehm2004d}. 
Radiative processes play an important role in comparisons with gamma-ray
data, and provide additional constrants on the model~\cite{Beacom,Sizun2006}.
In the past four years, the light dark matter model has been refined to account 
for the increasingly accurate data from INTEGRAL; 
the upper limit of the particles' mass has been placed 
anywhere from~$3$ to $20~\MeV$~\cite{Sizun2006,Ahn2005,Kasuya2006,Zhang2006}
and the theoretical aspects of this model have been studied considerably 
\cite{Boehm2004c,Fayet2006,Ascasibar2006}. In particular, 
the flux intensity measurements have allowed a fairly accurate determination 
of the cross-section of the annihilation reaction 
\cite{Ascasibar2006,Fayet2006b}.  Several authors 
have noted that increasingly accurate measurements of the morphology of the 
$511$~keV line would allow the mass of the proposed particle to be narrowed 
down even further \cite{Ahn2005,Kasuya2006,Rasera}.
\par
The annihilating dark matter scenario has attracted much attention because it is the 
easiest ``exotic" explanation to test experimentally. For instance, 
Ref.~\cite{Boehm2006a} focused on detecting a line from the galactic center that would 
result from the dark matter particles annihilating directly to photons, and 
Ref.~\cite{epem} proposed searches for the gauge boson involved in annihilation 
processes at high-energy $\epem$~colliders. We show that the light dark matter 
model can be tested cleanly and simply in the laboratory, using techniques
and methods that are readily available today.   Even taking into account the 
range of possible dark matter particle mass, the kinematical distinctions
between a dark matter production event and background events are clear enough
that a signal could be identified with a high degree of confidence.
Moreover, if dark matter were to be detected in such an experiment, 
the measurement of the cross-section would allow an independent 
check of the parameters in the light dark matter model, and by extension, provide 
significant information on the annihilations that take place at the center of 
the galaxy.

%%%%%%%%%%%%%%%%%%%%%%%%%%%%%%%%%%%%%%%%%%%%%%%%%%%%%%%%%%%%%%%%%%%%%%
%
\section{Resume of the Light Dark Matter Model}
\par
\begin{figure}
\begin{center}
\mbox{
\includegraphics[width=0.45\textwidth]{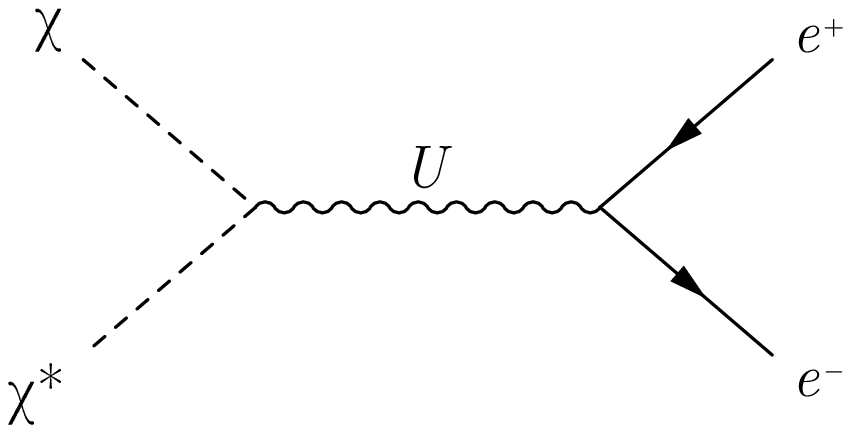}
\includegraphics[width=0.45\textwidth]{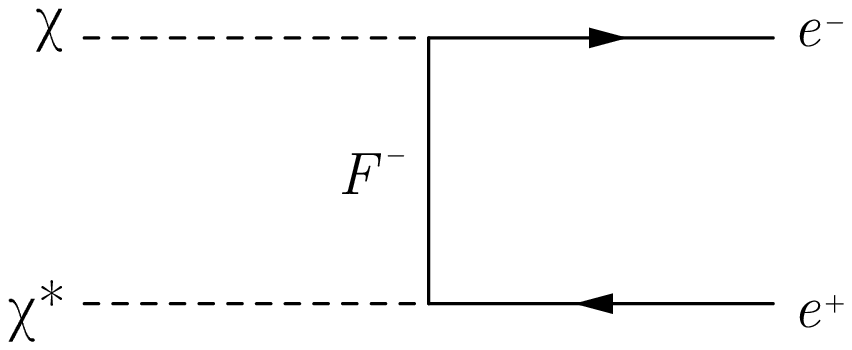}
}
\caption[.]{\label{F:dmannih}
\em dark matter annihilation into an $\epem$ pair,
through $U$-boson exchange in the $s$-channel, and 
$F^\pm$-fermion exchange in the $t$-channel}
\end{center}
\end{figure}

If the positron excess is explained by $\chi\chi^*\rightarrow\epem$,
then clearly there is an effective $\epem\chi\chi^*$ vertex.
Calculations by Boehm, Fayet and others have linked the strength
of this effective vertex to the observed dark matter abundance,
and the dark matter particle mass,~$\Mchi$.  
The $\epem\chi\chi^*$ interaction would be modeled by one or more 
intermediate particles.  According to a recent paper~\cite{Ascasibar2006},
one needs both a heavy fermion called the $F^\pm$ and a light neutral vector
boson called the $U$ to explain both the primordial abundance of
dark matter and the current rate of positron annihilation in the
galactic center.  In this model, the dark matter particles
are neutral scalars with masses on the order of $1$--$10~\MeV$.
There are other models in which $\chi$ is a fermion~\cite{FayetFermionChi},
but for concreteness, we will take $\chi$ to be a scalar.
The $U$-boson is also light, with $M_U > 2\Mchi$ and $M_U < 100~\MeV$,
and it decays mainly to $\chi\chi^*$.  
The mass of the $F^\pm$ fermion is at least several hundred~GeV
and plays no role in our process, so we do not consider it further.
Feynman diagrams for the dark matter annihilation are given in 
Fig.~\ref{F:dmannih}.
\par
The experiment that we propose does not depend crucially
on these assumptions.  Rather, this phenomenological model
serves as a guide for gauging the sensitivity of our experiment.
In order to estimate rates for our signal process, we must
specify the coupling constants.  We follow a recent paper by
Fayet~\cite{FayetLatest}, which links this model to several
experimental data, as well as a paper by Ascasibar,~{\em et~al.}~\cite{Ascasibar2006},
which links the model to various astrophysical data.  According to
Ascasibar~{\em et~al.}, the $U$~boson plays the dominant role
in fixing the primordial abundance of dark matter, while the $F^\pm$~fermion
controls the present day rate of annihilations, and hence the rate of 
positron production.  Fayet uses the dark matter abundance to 
constrain the $U$-boson couplings and does not hypothesize 
a heavy $F^\pm$ fermion.  He shows how particle physics data
constrain the $U$-boson couplings as a function of its mass.
Despite these differences, both papers
report similar constraints on the coupling constants of the $U$ boson 
to the dark matter particles, $\Cchi$, and to electrons, $\fe$.
Other publications placing similar constraints include
\cite{BoehmFayet,Fayet2006,Jacoby,Hooper}.
\par
\begin{figure}
\begin{center}
\includegraphics[width=0.95\textwidth]{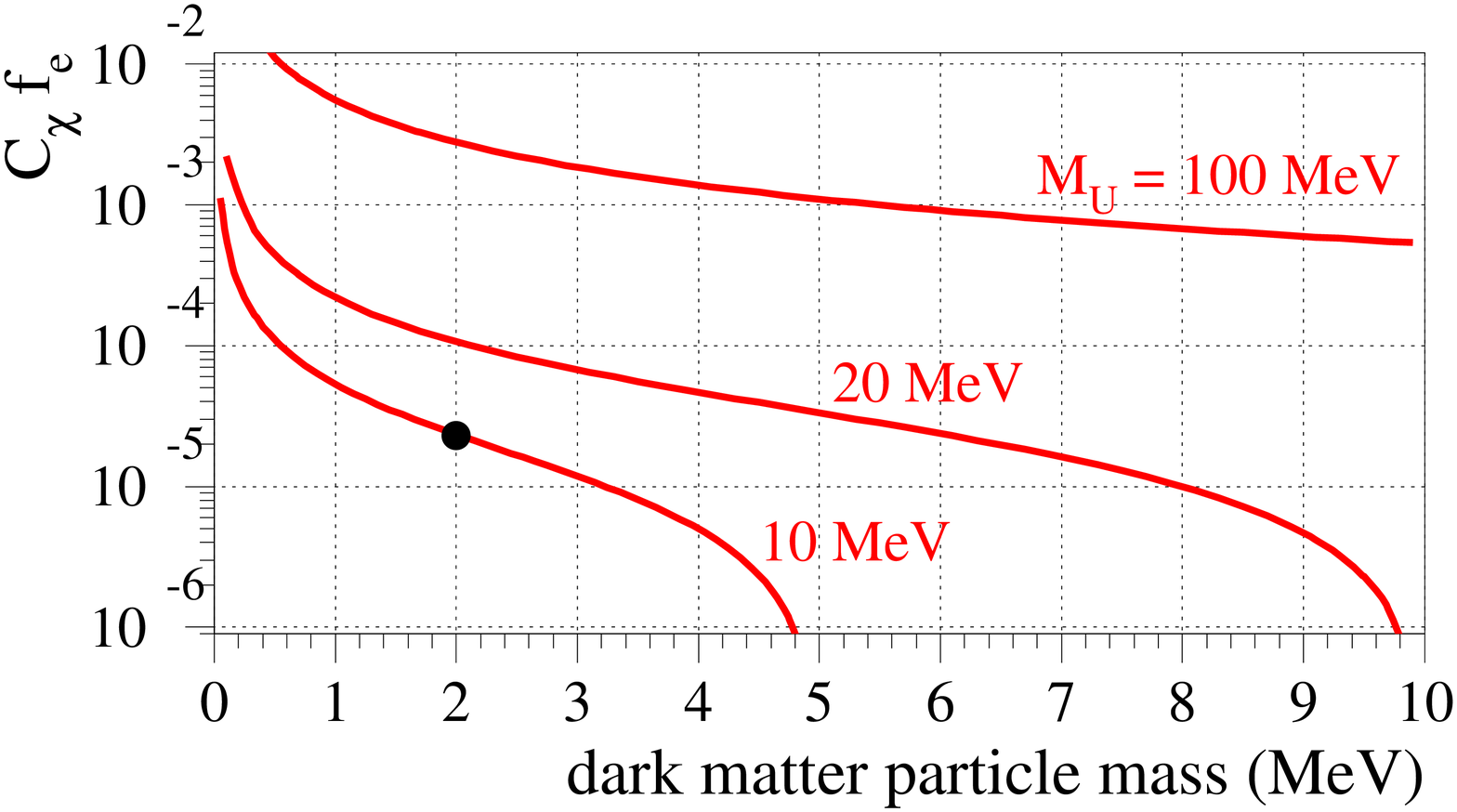} \\
\includegraphics[width=0.95\textwidth]{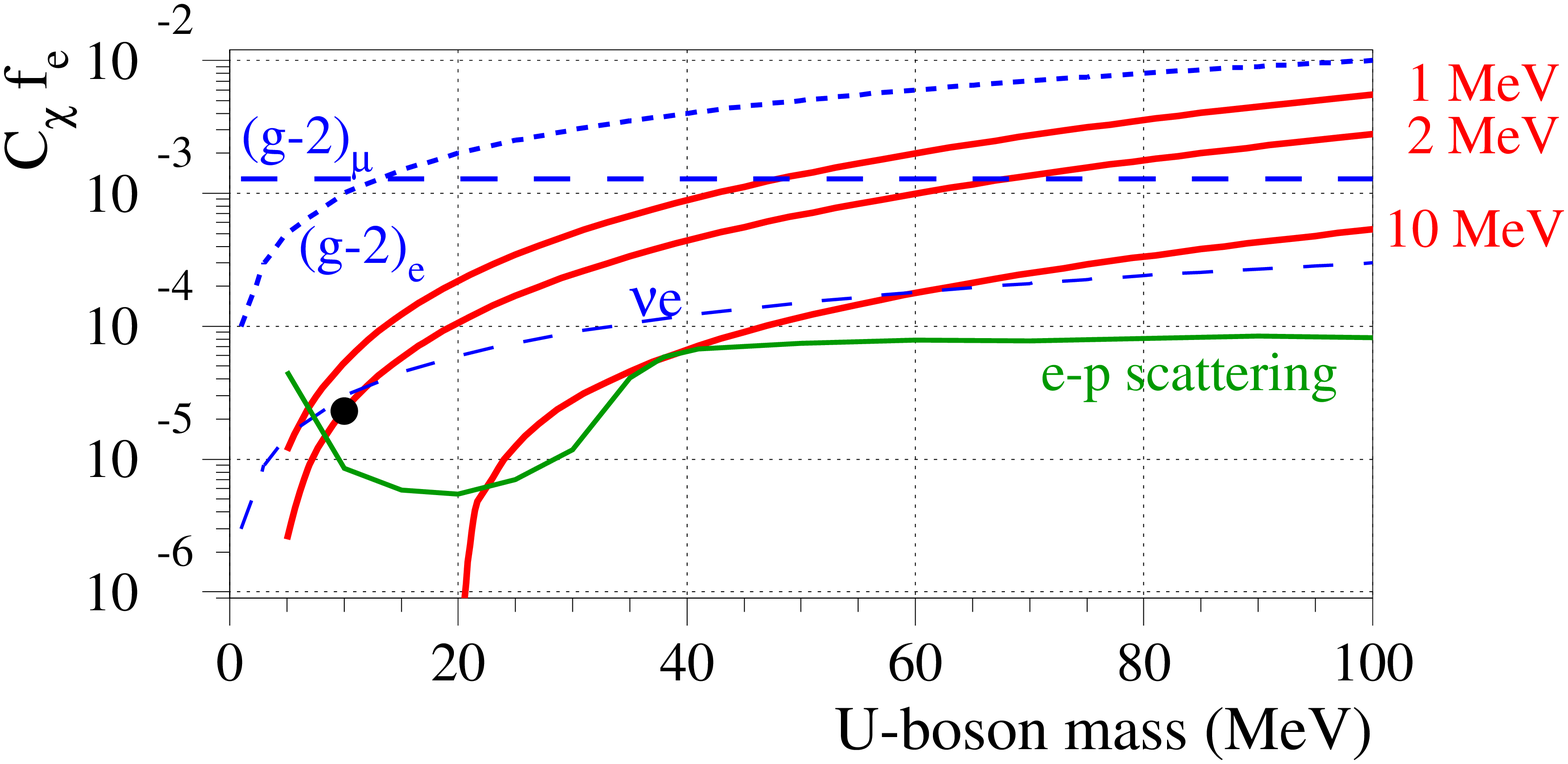}
\caption[.]{\label{F:constraints}
\em TOP: constraints on $\Cchi\fe$ as a function of $\Mchi$, for
three values of $\MU$.  BOTTOM:~heavy red lines show constraints on 
$\Cchi\fe$ as a function of $\MU$, for three values of~$\Mchi$.
The thinner dashed blue lines show constraints coming from the measurements
of $(g-2)_\mu$, $(g-2)_e$ and $\sigma(\nu e)$ (see text), 
from Ref.~\cite{FayetLatest}, and we have set $\Cchi = 1$.
The hook-shaped green curve indicates the sensitivity of the
first experimental design, discussed in Sec.~\ref{S:designI}.
The dots indicate our default choice for masses and couplings.
}
\end{center}
\end{figure}

If the dark matter abundance is used to constrain the $U$-boson couplings, then
\begin{equation}
\label{Eq:DMabundance}
  |\Cchi\fe| \approx 10^{-6}
  \frac{M_U^2 - 4M_\chi^2}{\Mchi\,(1.8~\MeV)} \, \sqrt{\Beeann}
\end{equation}
where $\Beeann$ is the fraction of all $\chi\chi^*$ annihilations which
result in an $\epem$ final state (see Eq.~(57) in Ref.~\cite{FayetLatest}).
We will assume that $\Beeann = 1$, although a lower value is possible if
there is a significant coupling of the $U$-boson to neutrinos.
If $\chi$ were a spin-$1/2$ fermion, then this formula would be modified
by a factor ${\cal{O}}(1)$.  Since our aim is to assess the viability of
an experiment, however, we will neglect any such factors in this paper.
Fig.~\ref{F:constraints}~(TOP) depicts this constraint as a function of~$\Mchi$
for three choices of~$\MU$.  The single dot shows our default choice of model
on which our rate estimates in Sec.~\ref{S:quasielastic} are based.  
Fig.~\ref{F:constraints}~(BOTTOM) shows
the constraint from Eq.~(\ref{Eq:DMabundance}) as a function of~$M_U$, for
three choices of~$\Mchi$.  It also shows the upper bounds on $|\Cchi\fe|$
derived by Fayet from three lepton-based experimental measurements, namely,
the measurement of $(g-2)_e$~\cite{(g-2)e}, 
the measurement of $(g-2)_\mu$~\cite{(g-2)mu}, 
and the $\nu-e$ cross-section measurement~\cite{nue}.
These constraints apply to the vector coupling of the $U$-boson to the electron;
more stringent constrains apply to the axial coupling.  We will set the
axial coupling to zero in our calculations.  The constraint from $\nu-e$
scattering assumes that $f_\nu = \fe$ and will be weaker if $f_\nu \ll \fe$.
The constraint from $(g-2)_\mu$ applies to our process only for $f_\mu = \fe$.
Fayet also derives constrains from $\psi$ and $\Upsilon$ decays, as well as 
atomic parity violation  experiments, but we will assume that the $U$-boson 
couples to leptons only.

%---------------------------
%---------------------------

%---------------------------
%---------------------------

\newpage

%
%%%%%%%%%%%%%%%%%%%%%%%%%%%%%%%%%%%%%%%%%%%%%%%%%%%%%%%%%%%%%%%%%%%%%%
%
\section{Low-Energy Quasi-Elastic $ep$-Scattering}
\label{S:quasielastic}
\par
We wish to exploit the different kinematic characteristics of
the signal process
$$
   e^- p \rightarrow e^- p \, U^{(*)} \rightarrow e^- p \, \chi\chi^*
$$
and of simple elastic scattering
$$
   e^- p \rightarrow e^- p 
$$
at low energy. The kinematics of the scattered electron and proton
are highly constrained for elastic scattering, and much less so
for the signal process.  Several kinematic distinctions can be made,
which allows an efficient and effective discrimination of the two
processes on the basis of kinematics alone.
\par
\begin{figure}
\begin{center}
\mbox{
\includegraphics[width=0.45\textwidth]{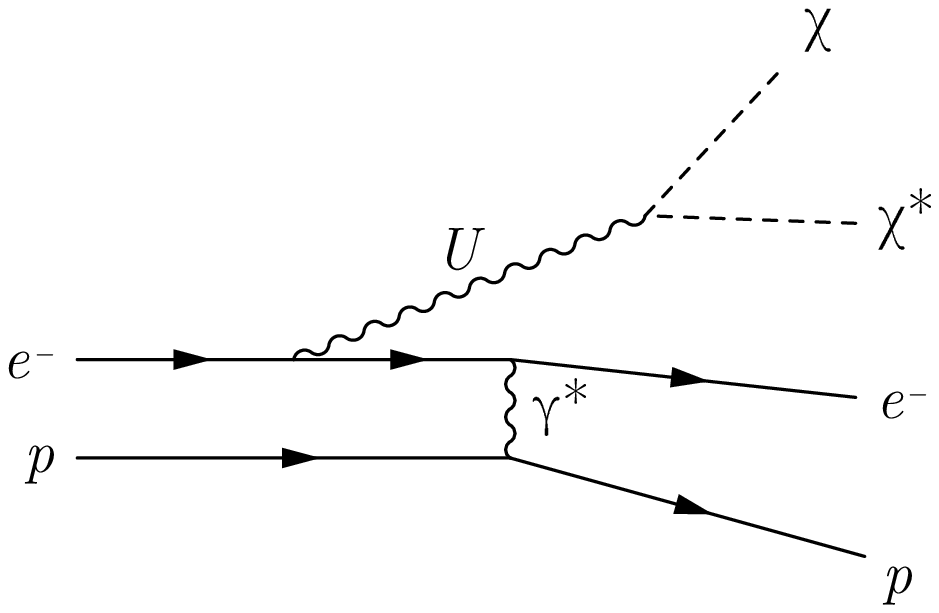}
\includegraphics[width=0.45\textwidth]{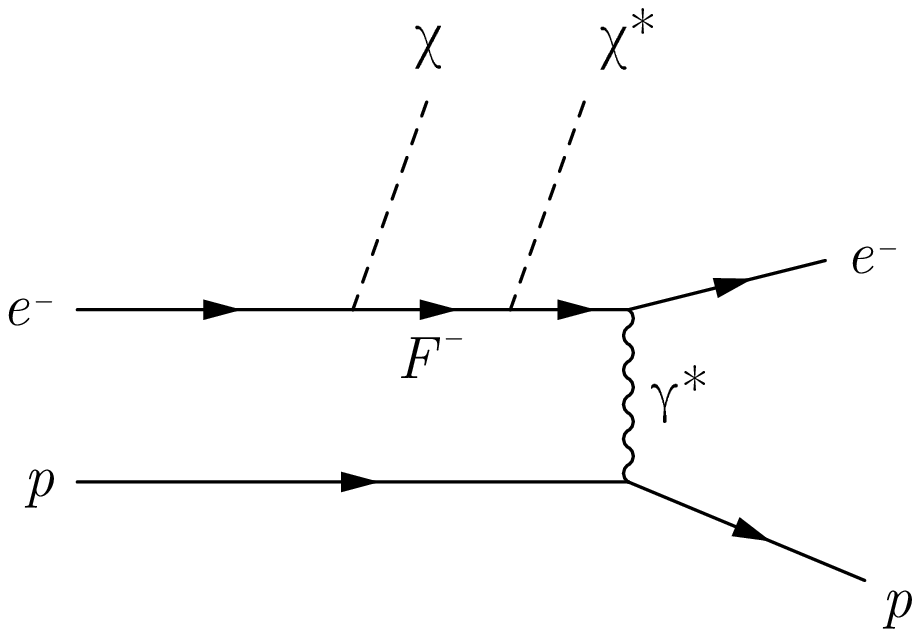}
}
\caption[.]{\label{F:signal}
\em example Feynman diagrams for the signal process.  We consider only
the $U$-boson diagrams in our estimates; the $F^\pm$-fermion
diagrams can be neglected.}
\end{center}
\end{figure}

Feynman diagrams for the signal process are given in 
Fig.~\ref{F:signal}.  It is important to note that the same set of
vertices appear in both our signal process and the dark matter annihilation
process (Fig.~\ref{F:dmannih}).   Consequently, our signal process {\em must} 
occur if this model is correct, and the signal cross-section can be related
directly to the dark-matter annihilation rates in the early universe
(which determines the dark matter relic density) and today (which
determines the strength of the $511$~keV gamma-ray line from the
galactic center).

%---------------------------
%---------------------------

\par
Since the dark matter particles are light, by hypothesis, we choose
to employ a low electron beam energy which essentially closes off 
all inelastic processes, leaving only elastic scattering as a
significant background process.  For example, we may take an
electron beam of energy $\Ebeam \approx 40~\MeV$ and a 
fixed hydrogen target, for which $\sqrt{s} \approx M_p + \Ebeam < M_p+\Mpi$,
since $\Ebeam \ll M_p$.  There can be no inelastic background
from ordinary strong-interaction processes, and if a pure hydrogen
target is used, there would be no nuclear excitations, either.
\par
A very small background will come from the process
$e^- p \rightarrow e^- p \nunubar$ mediated by an off-shell $Z$ boson.
The Feynman diagram is the same as the $U$-boson exchange diagram
(Fig.~\ref{F:signal}), with the $U$-boson replaced by a virtual $Z$-boson
and the scalar particles $\chi \chi^*$ replaced by $\nunubar$.
This background process is completely negligible -- about
nine orders of magnitude smaller than the hypothetical signal --
since $\MU \ll M_Z$.
\par
Higher-order QED radiative processes $e^- p \rightarrow e^- p \gamma (\gamma)$
also pose a background.  They occur at a smaller rate than elastic
scattering.  The final-state photons tend to emerge along the directions 
of the incoming and outgoing electrons, and hence do not change their 
directions~\cite{MoTsai}, although a small component of ``wide-angle''
Bremsstrahlung can occur and has been observed in muon-scattering 
experiments~\cite{EMCWAB}. Calculations for experiments at HERA and JLAB 
show extremely small rates outside a cone of $0.1$~rad~\cite{radcorr}.
In the case that the radiated photons emerge along the scattered electron 
direction, they can be detected together with the scattered electron.  
If they are emitted along the beam direction, thereby reducing the effective
$\sqrt{s}$ for the $ep$~interaction, the correlation of the electron and
proton directions changes very little.  This is a special feature of our
kinematical situation, in which $M_p \gg \Ebeam \gg M_e$.
Additionally, QED radiation peaks at low photon energies, which means
that the deviation of the scattered electron energy from the 
lowest-order results tends to be small, with no sharp kinematic
features.  The signal process peaks at large deviations, and there
are distinctive kinematic features due to the phase space needed
to produce two dark-matter particles, and/or an on-shell $U$-boson.
Finally, the QED radiation pattern varies very little with the beam
energy, while the signal process will have a strong dependence.
Hence, discrimination between the signal process and QED radiative 
processes will be possible if the signal rate is not too small.
\par
The concept for the experiment is the following: collide a well-defined
electron beam onto a hydrogen target, and observe the scattering angle
and energy of the outgoing electrons and protons.  For ordinary elastic
scattering, the energy of the scattered electron emerging at a given 
angle is constrained to a unique value.  So one would look for events 
with a significantly lower energy as evidence for the production of 
a final state that is {\em lighter} than a single pion.  Confirmation 
for an anomalous final state would come from the measurement of the 
scattering angle and energy of the outgoing proton.  The kinematic 
distributions for the final-state electron and proton would allow, 
in principle, a confirmation of the production of light invisible 
scalars of a particular mass, as opposed to the production of a pair 
of neutrinos.   In the absence of a signal, stringent limits could be 
placed on the masses and effective couplings of dark matter to electrons.
\par
The keys to identifying signal events are:
\be
\item
For a given scattering angle, the scattered electron will
have a much lower energy than in elastic scattering.
\item
The outgoing proton will be relatively slow;
for elastic scattering the proton is energetic.
\item
For a given electron scattering angle, the
scattering angle of the proton will vary over a wide range.
In elastic scattering, the proton emerges at a
unique angle.
\item
The electron and proton can be acoplanar for the signal
event, due to the momentum taken by the $\chi\chi^*$~pair;
for elastic scattering the electron and proton are
strictly back-to-back in the transverse plane, even if 
there is final-state radiation.
\item
The signal will increase rapidly as $\Ebeam$ is increased from threshold, 
while elastic scattering decreases as $1/E_{\mathrm{beam}}^2$.  There will
be a minimum beam energy corresponding to the threshold for producing
two dark-matter particles, below which there is no signal.
\item
The signal cross-section is less sharply peaked toward small electron
scattering angles.
\item
In principle, the $U$ boson may decay to an electron-positron
pair rather than the invisible $\chi\chi^*$ state.  The final
topology would contain {\em four} charged tracks.  If the $U$-boson
is on mass shell, then the mass distribution of the `extra' $\epem$ 
pair would give a peak at the $U$-boson mass.  If it is off mass
shell, then the mass distribution will be less sharply peaked
toward $2M_e$ than that due to photon conversions.
\ee
These very distinctive features allow an efficient selection
of events with very little background.  As described later in
this paper, a simple apparatus should allow signal events to
be identified at the rate of one event in a ten billion elastic
scatters, and a more sophisticated apparatus should achieve
a sensitivity better than $10^{-12}$ of the elastic scattering
rate.  Such experiments should easily cover the range of possible models
of this type, leading either to the discovery of the $U$-boson
or the exclusion of this and similar models.
\par
While we have based our calculations on a fairly specific model
and final state, our argument is not dependent on this model
in all of its details.  As already noted, the $\chi$ particle
might be a fermion instead of a scalar.  Furthermore, the
experiment we propose could be viewed as the direct production
of $U$-bosons, regardless of how they decay or whether they play
any role in dark matter phenomena.  In this case, the rates would
not depend at all on the coupling $\Cchi$, and the final states
might be dominated by other light particles.  If a signal
for an anomlous invisible final state were observed, then
follow-on experiments would be needed to confirm the connection
with dark matter (see Ref.~\cite{evacuated}).   In order to
provide a clear and consistent framework for our discussion,
however, we will follow fairly closely the light dark matter 
model described above.
\par
We proceed now to a discussion of the rates.

%------------------------------------------------------------
\subsection{Kinematics}
\label{S:kinematics}
\par
We remind the reader of the basic kinematics for elastic scattering
in order to frame our discussion and define our notation.
The target proton is effectively at rest in the laboratory frame with 
four-momentum $P_\mu = (M, 0, 0, 0)$, and the incoming beam electron is
highly relativistic with four-momentum $p_\mu = (E, 0, 0, p)$.
The outgoing electron has $p'_\mu = (E', p'\sin\theta, 0, p'\cos\theta)$
which defines the electron scattering angle~$\theta$.  The outgoing
proton has $P'_\mu$ and the four-vector of both dark-matter
particles we will write as~$W_\mu$.  Hence, $W_\mu = p_\mu + P_\mu
- p'_\mu - P'_\mu$, and $W_\mu = 0$ for elastic scattering.
For the signal process, $\min(W_{\mu} W^{\mu}) = 4\, M^2_\chi$.
\par
Following decades-old convention, we define the four-momentum
transferred $q_\mu = p'_\mu - p_\mu$ and $Q^2 = -q^2$.
To a very good approximation, $Q^2 = 4 E E' \sin^2\theta/2$.
The energy transfered in the lab frame is $\nu = E-E'$.
The kinematic condition for elastic scattering,
$P'^2 = P^2 = M^2$, implies $Q^2 \approx 2 M \nu$, after neglecting a
small term proportional to $m_e^2$.

%------------------------------------------------------------
\subsection{Elastic Cross-Section}
\par
The elastic electron-proton scattering cross-section, for
relativistic electrons is
\begin{equation}
\frac{d\sigma}{d\Omega} = \frac{\alpha^2 \hbar^2 c^2}{4E^2 \sin^4(\theta /2)}
\left ( \frac{E'}{E} \right ) [G_1(Q^2) \cos^2(\theta /2) + 2 \tau G_2(Q^2)
\sin^2(\theta /2)]
\label{E:elastic}
\end{equation}
where $\tau = Q^2 / 4M^2$ and
$$
G_1(Q^2) = \frac{G_E^2 + \tau G_M^2}{1+ \tau}, \hspace{3mm} G_2(Q^2) = G_M^2
$$
are functions of the proton electric form factor $G_E$ and
magnetic form factor $G_M$. At our low beam energies, $\tau \ll 1$, so we can
neglect the magnetic form factor. $G_E$ is approximated by the standard
dipole fit
$$
G_E(Q^2) \approx \left ( \frac{\beta^2}{\beta^2 + Q^2} \right )^2,
$$
where $\beta =$ 710 MeV. Under these conditions, Eq. (\ref{E:elastic})
simplifies to
$$
\frac{d\sigma}{d\Omega} = \frac{\alpha^2 \hbar^2 c^2}{4E^2 \sin^4(\theta /2)}
\left ( \frac{E'}{E} \right ) \left ( \frac{\beta^2}{\beta^2 + Q^2} \right )^4
\cos^2(\theta /2).
$$
$E'$ and $\theta$ are related by
$$
% \nu = E - E' = \frac{2E^2 \sin^2(\theta/2)}{M+2E \sin^2(\theta/2)}.
 E' = \frac{E}{1+(E/M)(1 - \cos\theta)}.
$$
The contribution of the form factor to the cross-section is on the order of
a few percent at low beam energies; this correction is necessary to get an
accurate estimate of the elastic background, but is not needed for an
order-of-magnitude estimate of the signal process.

\subsection{Signal Cross-Section}
\par
We investigated the signal process in two ways.  First, we performed a
semi-analytic calculation in which the matrix element was assumed to be
independent of all momenta.  Then we employed CompHEP~\cite{comphep}
and implemented the $U$-exchange process, an example of which is shown 
in the left-hand Feynman diagram of Fig.~\ref{F:signal}.

\subsubsection{Semi-Analytical Cross-section}
\label{S:semianalytical}
\par
To get a rough idea of the signal kinematics, we calculated
$d^2\sigma / d\Omega \,dE'$ assuming that the matrix element of the
signal process is independent of all momenta.
The phase-space calculation for a $2 \rightarrow 4$ process can be
rather involved, especially when several of the final-state particles
are massive, and finding an analytic expression for
$d^2\sigma / d\Omega \,dE'$ in terms of the four-vectors of all outgoing
particles is  not possible in general.  We simplified the calculation by
writing $W_{\mu}$ for the sum of the two four-vectors of the scalar particles
and expressing the cross-section in terms of the combined invariant mass
$W^2$ and the combined spatial velocity $\vec{W}$, treating these two
quantities as separate variables. In the calculations that follow, all
energies and momenta pertain to the center-of-mass frame.

\par
We begin with the Golden Rule for $2 \rightarrow 3$ scattering:
\begin{eqnarray*}
d\sigma &=& |\mathcal{M}|^2 \frac{1}{4\sqrt{(p \cdot P)^2 - (m_e M)^2}}
\left [ \left ( \frac{d^3 \vec{p'}}{(2\pi)^3 2E'} \right )
\left ( \frac{d^3 \vec{P'}}{(2\pi)^3 2E'_{p}} \right )
\left ( \frac{d^3 \vec{W}}{(2\pi)^3 2E_W} \right )
\right ] 
 \cr
 & & \times (2\pi)^4 \delta^4(p \ + P \ - p' \ - P' \ - W)
\end{eqnarray*}
where the relevant four-vectors are defined in Section~\ref{S:kinematics}.
Since we are only concerned with the shape of the phase space distribution, 
we will drop all numerical constants for the rest of this calculation. 
Making the standard approximation of a massless electron, we integrate out 
the proton momentum and the angular part of $\vec{W}$, using the 
delta function to obtain limits on $|\vec{W}|$:
\begin{equation}
|\vec{W}|_{\pm} = \frac{FE' \pm (\Etot - E')\sqrt{F^2 - 4W^2\Etot
(\Etot-2E')}}{2\Etot(\Etot-2E')},
\label{E:wlimits}
\end{equation}
where $\Etot$ is the total energy, $E'$ is the outgoing electron energy,
and $F \equiv \Etotsq + W^2 - 2\Etot E' - M^2$.
At this point
\begin{equation}
\frac{d^2\sigma}{d\Omega \, dE'} \propto
\int_{|\vec{W}|_{-}}^{|\vec{W}|_{+}} \frac{|\vec{W}|}{\sqrt{W^2 + |\vec{W}|^2}}
\,d|\vec{W}| = \sqrt{W^2 + |\vec{W}|_{+}^2} - \sqrt{W^2 + |\vec{W}|_{-}^2} .
\label{E:integral}
\end{equation}
Requiring that $|\vec{W}|_{\pm}$ be real, {\em i.e.,} that the term under the radical
in (\ref{E:wlimits}) be non-negative, gives the upper bound
\begin{displaymath}
(W^2)_{+} = M^2 + \Etotsq - 2\Etot E' - 2M\sqrt{\Etotsq - 2\Etot E'}
\end{displaymath}
with the lower bound $(W^2)_{-} = 4M_{\chi}^2$.  We use these bounds to integrate
the right-hand side of Eq.~(\ref{E:integral}) over $W^2$. It turns out that this 
integrand can be very well approximated by a square root function:
\begin{displaymath}
f(W^2) \equiv \sqrt{W^2 + |\vec{W}|_{+}^2} - \sqrt{W^2 + |\vec{W}|_{-}^2} \approx
K\sqrt{(W^2)_{+} - W^2}.
\end{displaymath}
The normalization constant $K$ is chosen so that the square root function matches
$f(W^2)$ at the endpoint $W^2 = (W^2)_-$; that is,
\begin{displaymath}
K = \frac{f((W^2)_-)}{\sqrt{(W^2)_+ - (W^2)_-}} \ ; \hspace{2mm}
f(W^2) \approx \frac{f((W^2)_-)}{\sqrt{(W^2)_+ - (W^2)_-}} \sqrt{(W^2)_+ - W^2}.
\end{displaymath}
(The other endpoint $W^2 = (W^2)_+$ is already taken care of since both the square 
root and $f(W^2)$ vanish there.) Performing the integration over $W^2$, we now have
\begin{eqnarray*}
\frac{d^2\sigma}{d\Omega\,dE'} & \propto &
\int_{(W^2)_-}^{(W^2)_+} \frac{f((W^2)_-)}{\sqrt{(W^2)_+ - (W^2)_-}} 
\sqrt{(W^2)_+ - W^2}\
 d\,(W^2) \\
& \propto & f((W^2)_-) \, ((W^2)_{+} - 4M_{\chi}^2) \\
& = & \left (\sqrt{4M_{\chi}^2 + |\vec{W}|_{+}^2} - \sqrt{4M_{\chi}^2 + |\vec{W}|_{-}^2}
\right ) ((W^2)_{+} - 4M_{\chi}^2)
\end{eqnarray*}
where $|\vec{W}|_{\pm}$ are evaluated at $(W^2)_- = 4M_{\chi}^2$.

A plot of this function, boosted from the CM frame to the lab frame, 
gives the broad phase-space curves in Fig.~\ref{F:escatt_comparisons}.
We note that this calculation is model-independent in the sense that
it makes no reference to the nature of the interaction.

\subsubsection{CompHEP Calculations}
\label{S:comphep}
\par
\begin{figure}
\begin{center}
\includegraphics[width=0.45\textwidth]{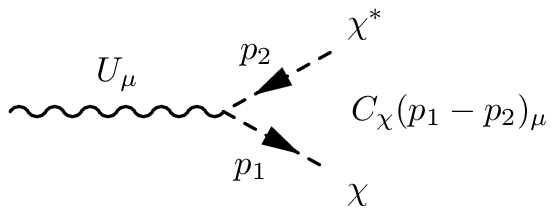}
\includegraphics[width=0.45\textwidth]{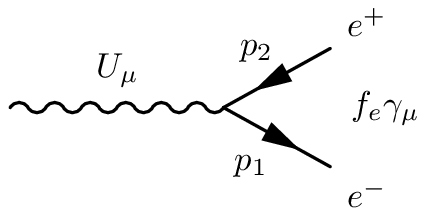}
\caption[.]{\label{F:Urules}
\em Feynman rules for the $U$-$\chi$ and $U$-$e$ vertices}
\end{center}
\end{figure}

We implemented $U$-boson exchange between electrons and dark matter particles.
An example Feynman diagram is shown on the left side of Fig.~\ref{F:signal}.
The proton was represented as a massive fermion with no internal structure, which,
for the very low beam energies we propose, is a very good approximation.
The mass of the dark matter particle was set to $\Mchi = 2~\MeV$, and the
mass of the $U$-boson was set to $\MU = 10~\MeV$.  Hence, the $U$-boson was
produced on mass shell. The coupling of the $U$-boson is assumed to be purely 
vectorial -- see Fig.~\ref{F:Urules} for the Feynman rules for the $U$-$\chi$
and $U$-$e$ vertices.  We imposed the relic abundance constraint, Eq.~(\ref{Eq:DMabundance}).
Taking $\Cchi = 1$ and $\Beeann = 1$, this gives $\fe \approx 2.3\times 10^{-5}$.

%---------------------------
%---------------------------

\par
For the purposes of this paper, we set the beam energy to $\Ebeam = 40~\MeV$,
and we considered a narrow range of electron scattering angle
$89.4^\circ < \theta_e < 90.6^\circ$, which follows from the
conceptual design presented in Section~\ref{S:designI}.
\par
\begin{figure}
\begin{center}
\includegraphics[width=0.85\textwidth]{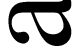} \\
\includegraphics[width=0.85\textwidth]{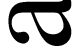} \\
\caption[.]{\label{F:dsigma}
\em signal cross-sections.  TOP:~accepted cross-section as a
function of~$\MU$.  The solid line shows the result with a fixed
coupling constant, and the dashed line shows the result when the
constant varies with mass according to Eq.~(\ref{Eq:DMabundance}).
Here, ``accepted'' refers to a limited angular range for the 
scattered electron: $89.4^\circ < \theta < 90.6^\circ$.
BOTTOM:~cross-section as a function of the beam energy, for three
values of~$\MU$, as indicated.  The curves for $\MU = 5~\MeV$ and
$20~\MeV$ have been multiplied by factors of~$20$ and~$0.2$,
respectively.}
\end{center}
\end{figure}

The mass of the $U$-boson is not well constrained.
If the product of coupling constants $\Cchi\fe$ is held fixed, then the
signal cross-section drops rapidly as a function of~$\MU$, as depicted
by the solid curve in Fig.~\ref{F:dsigma}~(TOP).  The presence of the
threshold at $\MU \approx 40~\MeV$ is evident.  If the $M_U$ boson were
heavier, then one would run this experiment at a higher beam energy.
Backgrounds will not increase so long as $\Ebeam < 135~\MeV$, and in
fact the signal cross-section will increase, up to a certain point. According to 
Eq.~(\ref{Eq:DMabundance}), the constraints $\Cchi\fe$ should be adjusted
as a function of~$\MU$.  The result for the cross-section is shown as the 
dashed curve, which is much flatter than the solid curve, showing that
the rate of this process is indeed tied directly to the rate of dark matter 
annihilation.  Fig.~\ref{F:dsigma}~(BOTTOM) shows the variation of the signal 
cross-section as a function of beam energy, for three choices
of~$\MU$.   A typical threshold behavior is evident.

%---------------------------
%---------------------------

%---------------------------
\begin{figure}
\begin{center}
\includegraphics[width=0.85\textwidth]{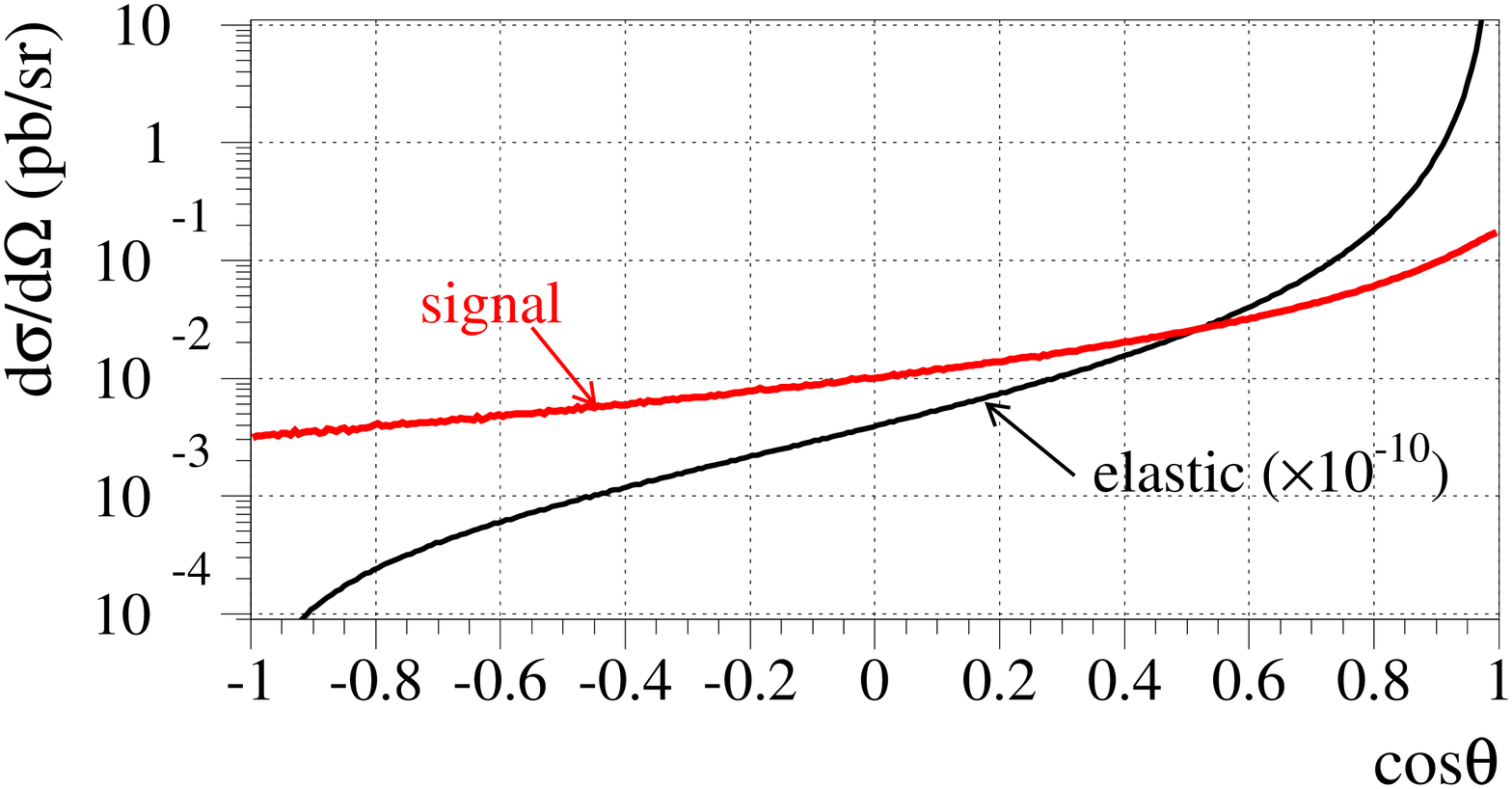} \\
\includegraphics[width=0.85\textwidth]{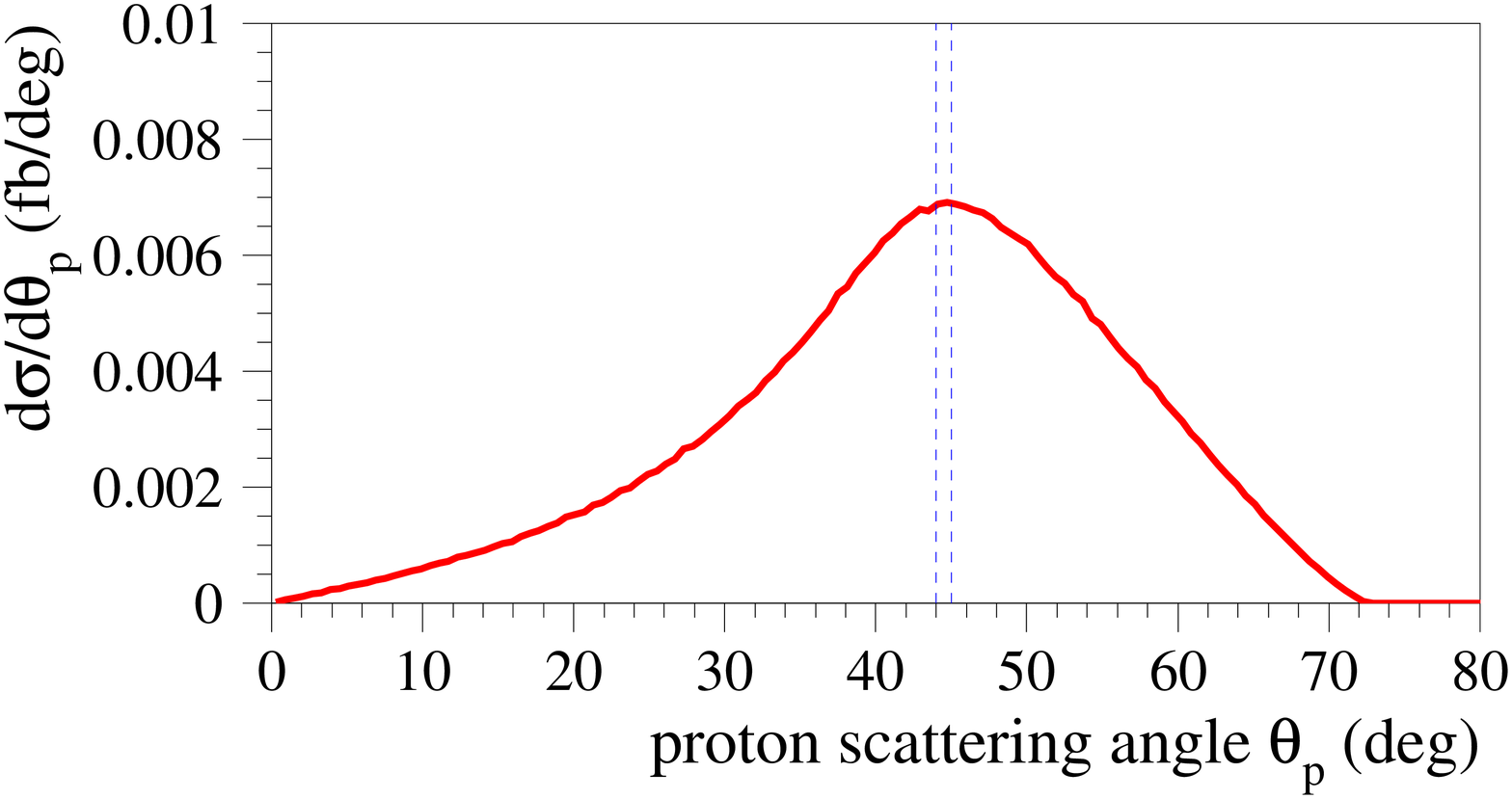} \\
\caption[.]{\label{F:dsigma2}
\em differential signal cross-sections as a function of angles.
TOP:~$d\sigma/d\Omega$ plotted as a function of $\cos\theta$.
The elastic scattering cross-section has been scaled down by a
factor of~$10^{-10}$.
BOTTOM:~$d\sigma/d\theta_p$ showing the broad distribution expected
for the signal.  The vertical dashed lines show the very narrow range 
expected for elastic scattering even allowing for radiative processes.}
\end{center}
\end{figure}
%---------------------------

\par
Since the signal process is inelastic, the electron is scattered at
larger angles than for the background process -- see
Fig.~\ref{F:dsigma2}~(TOP).  Clearly the ratio of signal to background
is a strong function of the scattering angle, and so we have chosen
$\theta = 90^\circ$, as an example.  An optimized choice of $\theta$ would 
depend on $\MU$ as well as the details of the apparatus, which is beyond
the scope of this paper.
\par
The wide range of proton scattering angles,~$\theta_p$, is also a
distinctive feature of the signal process. Fig.~\ref{F:dsigma2}~(BOTTOM)
shows $d\sigma/d\theta_p$, with the electron scattering angle~$\theta_e$ 
constrained as above.  The dashed lines indicate the narrow range of~$\theta_p$
expected for elastic scattering.  This range does not change even when
the incoming electron emits an energetic photon during the scattering process.

\par
\begin{figure}
\begin{center}
\includegraphics[width=0.75\textwidth]{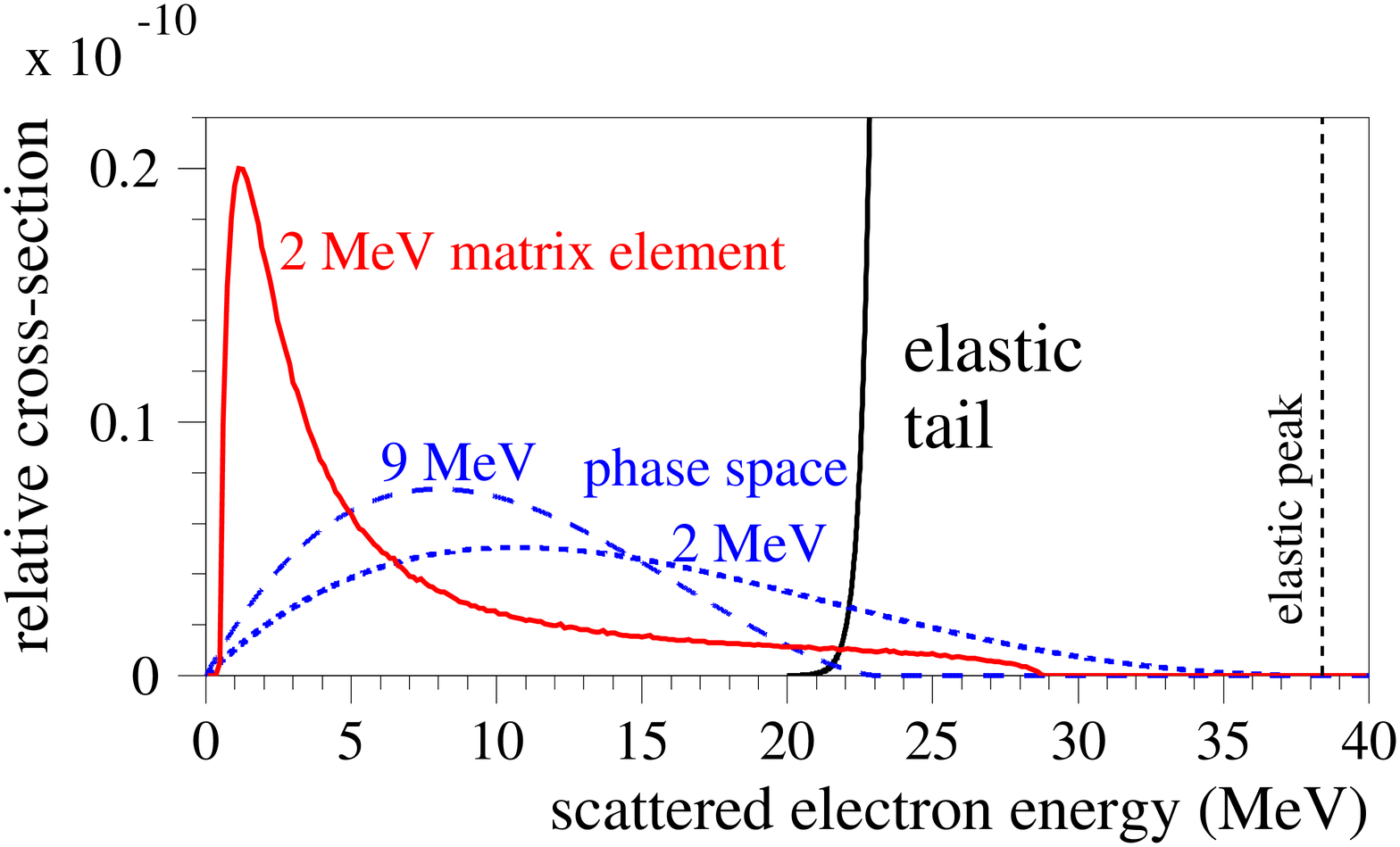}
\caption[.]{\label{F:escatt_comparisons}
\em signal shapes compared to a Gaussian centered at $E' = 38.4~\MeV$ 
with an r.m.s. of~$6\%$ (see text).
Two examples of phase-space distributions are shown, for
$\Mchi = 2~\MeV$ and~$9~\MeV$ (see Sec.~\ref{S:semianalytical}).
One example of a CompHEP matrix-element
calculation is shown, for $\Mchi = 2~\MeV$ and $\MU = 10~\MeV$
(see Sec.~\ref{S:comphep}).}
\end{center}
\end{figure}

An important feature of the signal process is the reduced energy
of the outgoing electron.  For $\Ebeam = 40~\MeV$ and $\theta = 90^\circ$,
perfect elastic scattering gives $E' = 38.4~\MeV$.  For the signal process,
$E'$ follows a much broader distribution with a peak at rather low values, 
as shown in Fig.~\ref{F:escatt_comparisons}.  It is an interesting feature
of quantum field theory that a massless spin-$1$ boson, such as the photon,
is emitted with an energy that peaks toward the minimum possible value,
while a massive spin-$1$ boson, such as the $U$-boson considered here,
is emitted with a momentum that peaks toward the maximum possible value.
Thus the CompHEP calculation, based on the matrix element for $U$-boson
exchange, shows a scattered electron energy distribution which peaks
toward small values.  The phase-space calculation, described in 
Section~\ref{S:semianalytical}, sets the matrix element to a constant, 
resulting in a broader distribution for the scattered electron energy
as shown in Fig.~\ref{F:escatt_comparisons}.   
\par
On the basis of these CompHEP calculations, and for 
$89.4^\circ < \theta_e < 90.6^\circ$, the accepted signal cross-section is 
$2.13\times 10^{-4}~\pb$, and the cross-section for elastic scattering
is $8.0\times 10^{5}~\pb$, which gives a ratio of cross-sections of 
$2.7\times 10^{-10}$.  We  checked these results using CalcHEP~\cite{calchep}.
In the next section we describe experiments which should
be able to probe this range successfully, and either
establish the existence of this signal process
or rule it out definitively.

%---------------------------
%---------------------------

\newpage
%
%%%%%%%%%%%%%%%%%%%%%%%%%%%%%%%%%%%%%%%%%%%%%%%%%%%%%%%%%%%%%%%%%%%%%%
%
\section{Proposed Experiments}
\par
The two proposed experiments bring a low-energy electron beam  ($\Ebeam = 40~\MeV$)
onto a fixed hydrogen target, and record a large number of elastic scattering 
events.  Each design exploits kinematic differences as discussed in 
Section~{\ref{S:quasielastic}} above.
The first design is simple and is meant to demonstrate
the concept, while the second is somewhat more ambitious,
though easily within the capabilities of existing techniques.

\subsection{Experimental Design I}
\label{S:designI}
\par
This design exploits the correlations between the scattering angle and 
outgoing electron energy, and the proton angle and velocity.  Fixing the
electron scattering angle in a narrow range $89.4^\circ < \theta < 90.6^\circ$,
the other three quantities should be centered in narrow ranges around
certain values, namely, scattered electron energy $E' = 38.4~\MeV$, 
scattered proton momentum $P'_p = 55.4~\MeV$, and angle
$\theta_p = 43.8^\circ$.  We imagine that the electron is detected
in a high-resolution electromagnetic calorimeter placed at $\theta = 90^\circ$
and covering $2\pi$ in azimuth, 
and the proton is detected in a Silicon strip detector placed at
$\theta_p = 43.8^\circ$.  Elastic scattering would give a coincidence
in the calorimeter and Silicon detectors.  The calorimeter would
measure the energy of the electron, and the silicon would measure
the kinetic energy of the proton due to its very high $dE/dX$, since
$v \sim 10^{-3} c$ for the proton. 
\par
\begin{figure}
\begin{center}
\includegraphics[width=0.75\textwidth]{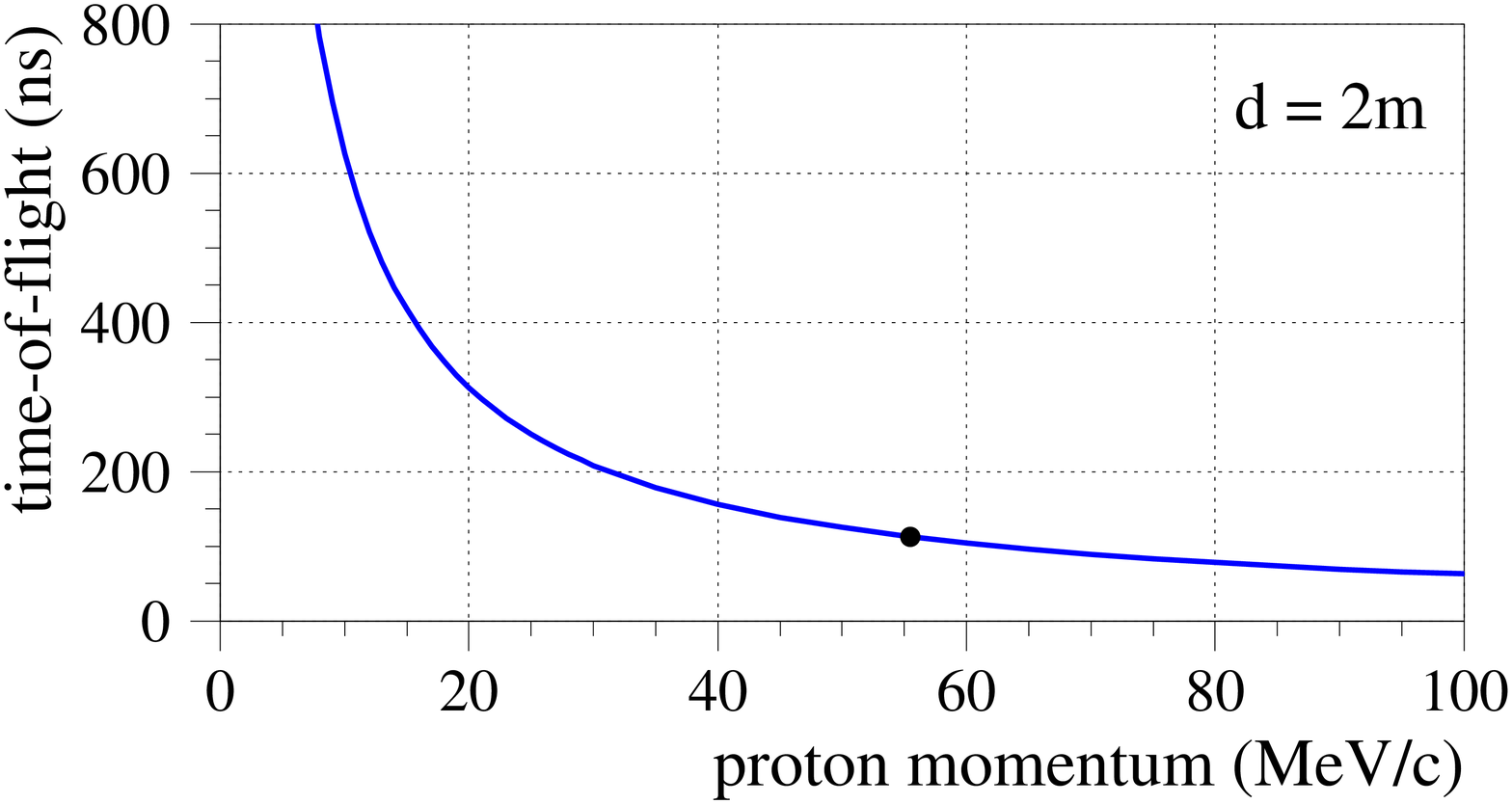}
\caption[.]{\label{F:TOF}
\em time-of-flight for the outgoing proton, when it
traverses a distance of two meters.  The dot marks the
expected momentum for elastic scattering in experimental
design~I.}
\end{center}
\end{figure}

A redundant measurement of the proton velocity would come from measuring 
its time-of-flight across a fixed distance.  We present this time-of-flight 
as a function of the proton momentum in Fig.~\ref{F:TOF}, where the dot 
indicates the expected value $P'_p = 55.4~\MeV$ for elastic scattering.  
The resolution on the proton momentum worsens as the momentum increases.
Nonetheless, for a distance $d = 2$~m, and for a time-of-flight resolution
of $2$~ns, the measurement uncertainty on $P'_p$ would be only $1~\MeV$,
which is more than adequate for these purposes.
\par
The energy resolution of the calorimeter is of paramount importance.
Recent examples of high-resolution electromagnetic calorimetry
based on  high-purity single crystals suggest that a resolution
of about $6\%$ could be achieved for $E' = 38.4~\MeV$~\cite{emcal}.
Fig.~\ref{F:escatt_comparisons} compares a purely Gaussian peak
centered on this value for~$E'$ with an r.m.s. of~$6\%$ to
three signal distributions generated according to the
phase space model, and the matrix-element calculation.  
The Gaussian curve is normalized to unity and the three 
signal shapes are normalized to~$10^{-10}$.
\par
A clear separation between signal and background from the elastic
peak is evident, and in this simple picture it is easy in principle to
isolate a signal by requiring $E' < 20~\MeV$, for example.
This cut would retain more than 75\% of the signal,
while the Gaussian is effectively removed.
\par
\begin{figure}
\begin{center}
\includegraphics[width=0.85\textwidth]{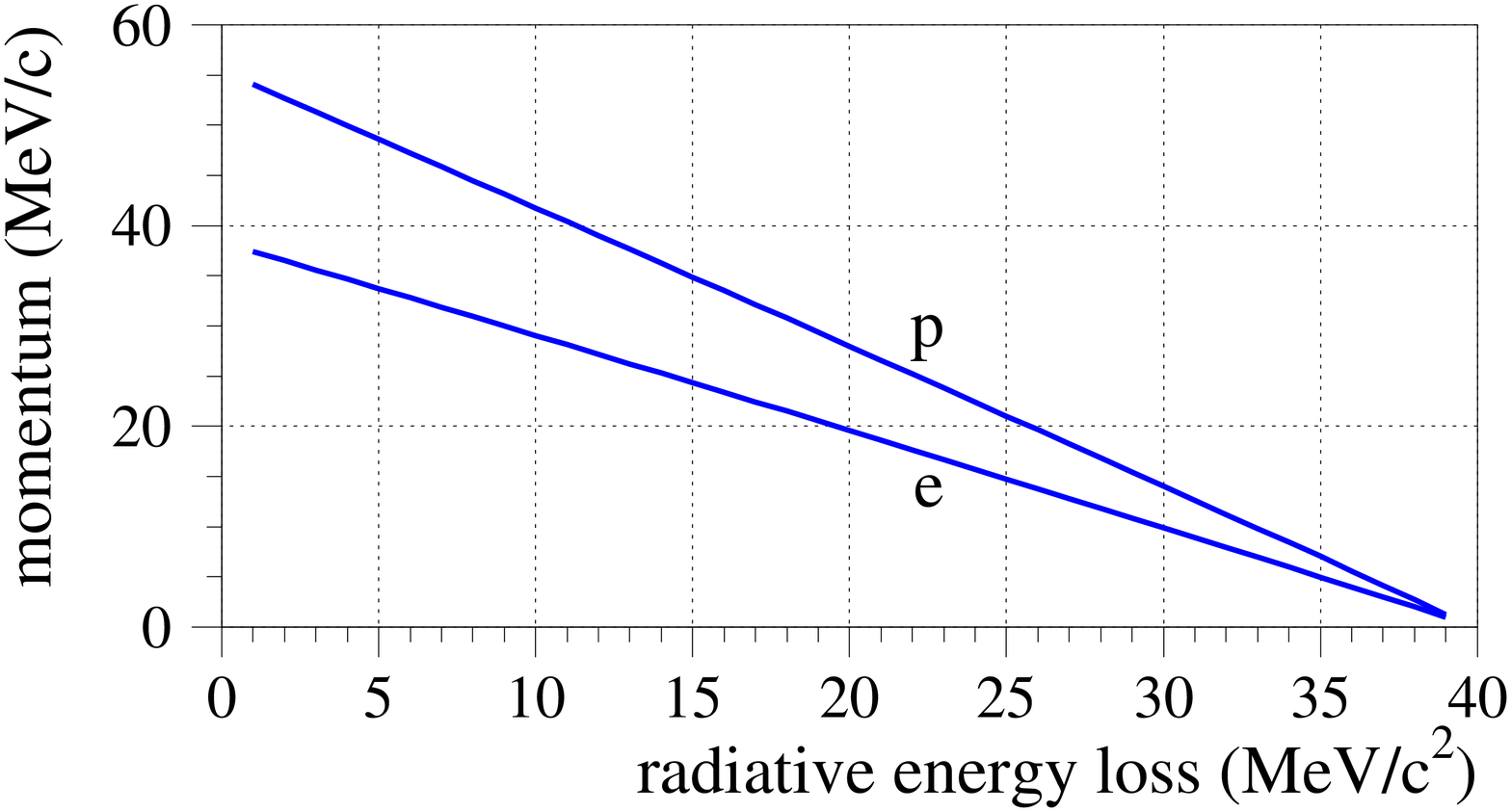}\\
\includegraphics[width=0.85\textwidth]{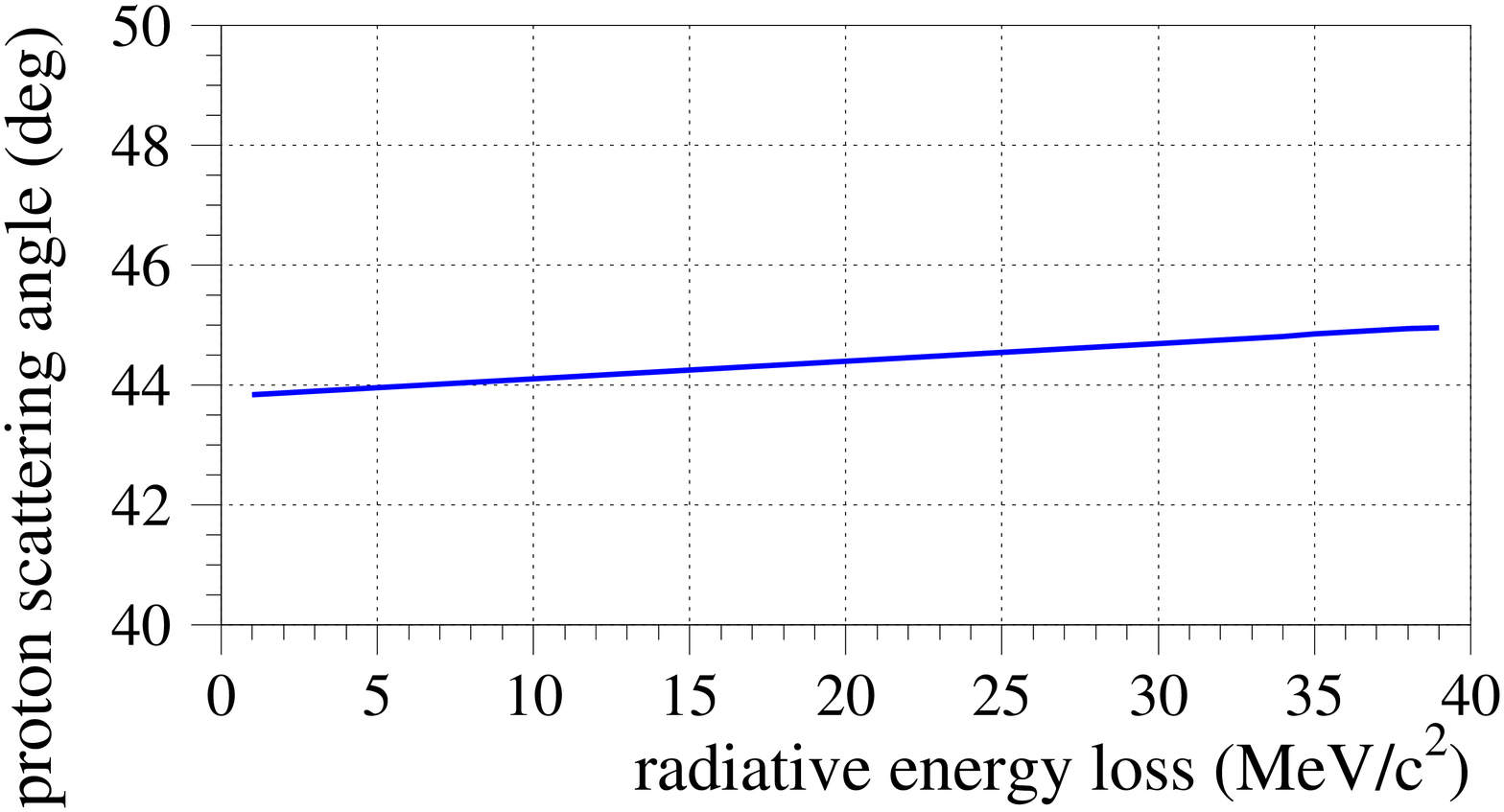}
\caption[.]{\label{F:radloss}
\em the impact of radiative energy loss due to photons emitted by the incoming
electron on the kinematics of elastic scattering.  TOP:~the momentum of the 
outgoing particles.  BOTTOM:~the variation of the proton scattering angle~$\theta_p$.}
\end{center}
\end{figure}

The signal from elastic scattering will not correspond exactly
to a Gaussian, however, due to resolution effects in the calorimeter
and due to radiative processes mentioned in  Section~\ref{S:quasielastic}.  
It is beyond the scope of this paper to try to estimate the exact 
shape of the elastic scattering peak.  One should note that
many of the  radiated photons will be emitted along the direction
of the outgoing electron, and hence will strike the same
calorimeter element that the scattered electron does.
This element will therefore register the energy radiated
by the electron together with the photon, providing a kind
of automatic correction for the radiation.  If photons are emitted
by the incoming electron resulting in an effective radiative
energy loss for the interaction, the energy of the outgoing electron
and proton decrease proportionally, as shown in the top plot of
Fig.~\ref{F:radloss}.   However, the angle of the outgoing proton 
hardly changes at all, as shown in the bottom plot of Fig.~\ref{F:radloss}.
Consequently, a robust signal for elastic scattering is available.
Radiation emitted at wide angles is extremely rare, perhaps 
on the order of $10^{-6}$~\cite{radcorr}, and in principle could be tagged.
\par
Scattering from residual gas molecules which have migrated far from the
nominal interaction point might mimic the kinematics of the signal,
and thereby constitute a background that is not easy to reject on the
basis of our kinematic cuts.  An assessment of the level of this 
background requires a detailed design and perhaps some level of
proto-typing, which is beyond the scope of this paper.
\par
It is difficult to estimate the power of this apparatus to reject
elastic scattering.  Nonetheless, it may be plausible that 
our signal could be isolated from elastic scattering.
The signature would be an outgoing electron at low $E'$ values 
(well below $25~\MeV$) and either the absence of a proton, or a slow
proton exiting at an angle significantly different from~$44^\circ$
(see bottom plot of Fig.~\ref{F:dsigma}), and with an azimuthal
angle with respect to the electron different from~$180^\circ$.
QED radiative processes would not give a peak at low~$E'$, rather, 
they would produce a tail to the Gaussian peak shown in 
Fig.~\ref{F:escatt_comparisons}.   
The variation of any signal with $\Ebeam$ and $\theta$ is quite
different from the variation of the background 
(see Fig.~\ref{F:dsigma}~(BOTTOM) and Fig.~\ref{F:dsigma2}~(TOP)).

%---------------------------
%---------------------------

%---------------------------
%---------------------------

\par
\begin{figure}
\begin{center}
\includegraphics[width=0.65\textwidth]{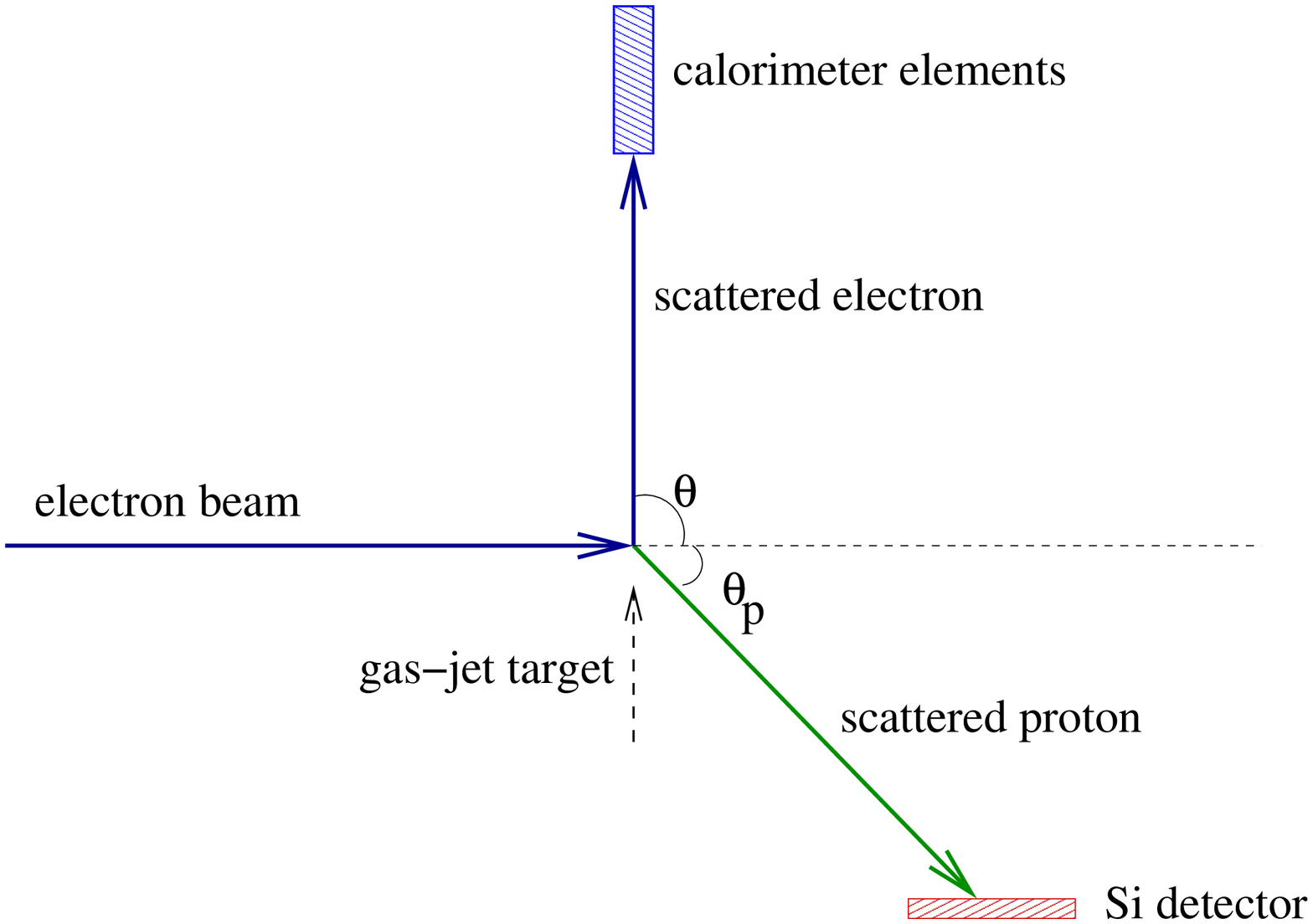}
\caption[.]{\label{F:designI}
\em experimental design I}
\end{center}
\end{figure}

As a toy example, we imagine an incoming beam with $E = 40~\MeV$
and a current of~10~mA.  It impinges on a gas-jet hydrogen target
with a density of $\rho = 2 \times 10^{18}$~atoms/cm$^2$~\cite{gasjet}.
Our calorimeter consists of one ring of elements arrayed at $\theta  = 90^\circ$,
as shown in Fig.~\ref{F:designI}.  The calorimeter elements are single
high-purity PbWO$_4$ crystals shaped as regular polyhedra which project
toward the target.  The front face might be 2~cm~$\times$~2~cm, and
the elements might be placed~$1$~m from the target. We assume a Gaussian 
resolution of 6\%, or $2.4~\MeV$ at $38~\MeV$~\cite{emcal}.
Additional plastic scintillator would be used to indicate when a shower
has leaked laterally outside the crystals, and when a wide-angle
photon has been emitted.  
The proton is detected with a silicon strip detector placed at the
angle expected for elastic scattering.  Additional detectors
could be placed at other angles to confirm a proton scattered
in the signal process. The experimental region would be
filled with a low-$Z$ material, such as Helium gas, to minimize
multiple scattering and energy loss of the proton.

%---------------------------
%---------------------------

\par
We integrated Eq.~\ref{E:elastic} numerically to obtain an accepted
cross-section for this simple apparatus, and obtained $0.8~\mu$b.
With the stated beam current and target density, the rate of elastic
scatters is $2\times 10^5$~events/s.  With two months' good running 
time (10 hours per day for fifty days),
one would accumulate on the order of $10^{11}$ elastic scattering events.
\par
An accurate estimate of the sensitivity of this experiment to our
signal process would require detailed simulations
of the apparatus, radiative and other energy loss processes.
We cannot provide such an estimate at this time.
However, a rough order-of-magnitude estimate can be made as follows.
Fig.~\ref{F:escatt_comparisons} shows a very clear kinematic 
distinction based on the measured scattered electron energy.
A simple cut such as $E' < 20~\MeV$ amounts to $8\sigma$ below
the elastic peak.  For an ideal calorimeter, none of the $10^{11}$ elastic 
scatters would survive this, while $> 75\%$ of the signal would survive,
depending on~$M_\chi$.   In practice, this simple cut on $E'$ would
be reinforced by the powerful cuts on the outgoing proton.
A positive signal of $15$~events would correspond to a signal 
cross-section that is a about factor of~$2\times 10^{-10}$ smaller 
than the elastic scattering cross-section, or about $0.2$~fb.  
The absence of any signal would correspond to an upper limit of about 
$3\times 10^{-5}$~pb, for our default choice of coupling constants
and masses.  We can estimate how such an upper limit would vary
as a function of~$\MU$, and the result is shown as the hook-shaped
curve in Fig.~\ref{F:dsigma}.  Apparently, this simple apparatus 
has a good chance to observe a signal or rule out this model.
\par
If a signal were observed in the energy spectrum of the scattered electron,
then one could expand the coverage of the Silicon detector so that the
recoil proton would be well measured for signal events.  Then the kinematics
of the initial state and the outgoing electron and proton would be known 
on an event-by-event basis, and it would be possible to measure the mass 
of the invisible final state by computing the recoil mass.  
Specifically, with our definition of the missing four-momentum, $W_\mu = p_\mu - p'_\mu 
+ P_\mu - P'_\mu$, the invariant mass of the two dark-matter particles
would be $M_{\chi\chi}^2 = W_\mu W^\mu$.  There will be a threshold 
of $2\Mchi$ for $\sqrt{W_{\mu} W^{\mu}}$, and if the $U$-boson is produced
on mass shell, then there will be a peak at $\sqrt{W_{\mu} W^{\mu}} = \MU$.
A detailed simulation is needed in order to understand the resolution
of the recoil mass.

\subsection{Experimental Design II}
\par
This design augments the calorimetry of the first design with
an open-geometry tracking device and a magnetic field.
Precise measurements of the momenta and direction of the outgoing
electron and proton would provide powerful handles against 
background processes.  As pointed out in Section~\ref{S:quasielastic},
there are angular variables which can be exploited to identify a signal.
Furthermore, such measurements serve to verify the electron energy
as measured by the calorimeter and thereby reduce radiative backgrounds.
\par
\begin{figure}
\begin{center}
\includegraphics[width=0.65\textwidth]{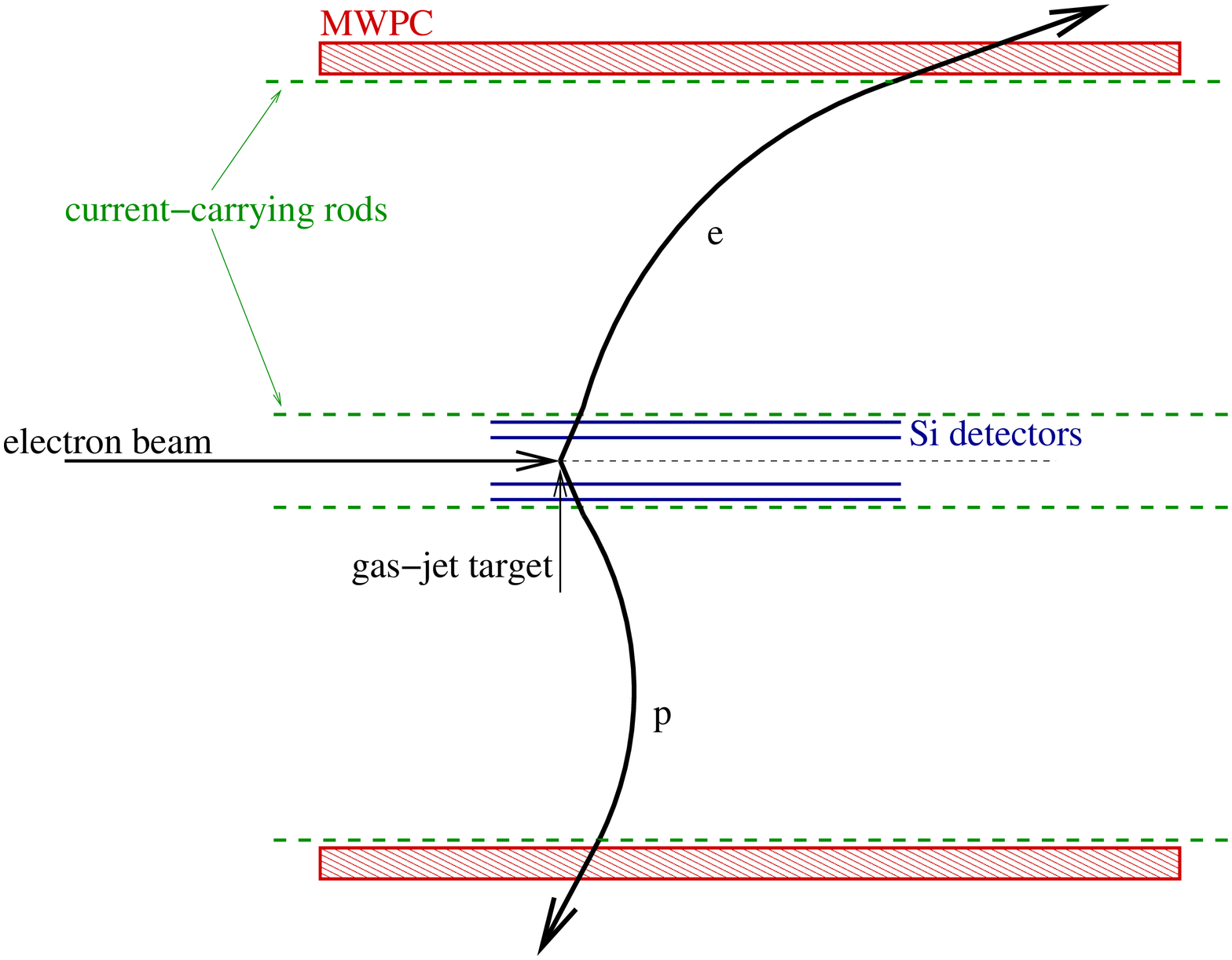}
\caption[.]{\label{F:designII}
\em experimental design II}
\end{center}
\end{figure}

We envision a cylindrical geometry coaxial with the beam --
see Fig.~\ref{F:designII}.
The region around the target would be instrumented with two layers
of precision silicon detectors which would provide the position and
angles of the outgoing electron and proton.  For the sake of discussion, we
take the radii of these two cylinders to be $1.5$~cm and $2.5$~cm.  
It is likely that the inner cylinder could be reduced to a few~mm, 
depending on the properties of the beam.  Directly outside the second 
cylinder we generate a {\em toroidal} magnetic field.  A small number 
of conducting rods will run parallel to the beam and carry a modest current.
A circumferential magnetic field is generated with a strength
on the order of hundreds of Gauss.  This field bends the electron and
proton but they remain co-planar with the beam, which is important
for triggering and data analysis.  At a radius of 1~m, we install
a cathode strip or multi-wire proportional chamber to measure the
$z$-coordinates of the electron and proton.  This measurement together
with the initial trajectory measured by the silicon detectors gives
us the momenta of the two particles.  Another set of conducting rods
carries a current which terminates the magnetic field.  Finally, an 
extended version of the calorimeter measures the energy of the 
scattered electron, independently of the tracking system, as well
as any photons radiated away from the beam.
\par
All of the proposed detector elements are ordinary, and many examples
exist in working detectors.  The scale of the apparatus is modest,
as is the number of instrumented channels.  The event rate would
be rather low compared to modern high-energy physics experiments,
so read-out electronics would not need to be especially fast.

%---------------------------
%---------------------------

\subsubsection{Analysis Plan}
\par
The measurements $z_0, z_1, z_2$ of the electron position at the surfaces of 
the three nested cylinders allow an accurate identification of a dark matter 
production event. The accuracy of this analysis plan depends significantly on 
the properties of the silicon detectors used at the surfaces of 
the inner two cylinders; in particular, there is an absolute ``worst-case'' 
error $\Delta z$ in the measurement of position, on the order of $100~\mu$m. 
In the following discussion we suppose that the incoming electron beam is 
traveling along the $z$-axis in the positive direction. For a given event, we 
compute the measured scattering angle $\theta_{\mathrm{meas}}$ 
from $z_0, z_1$, and the distance $R_1 - R_0$ between the two inner 
cylinders. We then simulate the path of the electron \emph{assuming} the 
event was an elastic collision, obtaining the outgoing electron energy from 
the kinematics of the reaction, and find the expected position $z'_2$ at the 
outer detector wires. At this point all events with $z_2 < z'_2$ are rejected.

If the event were in fact an elastic collision, and 
$\theta_{\mathrm{meas}} > \theta_{\mathrm{true}}$, then $z_2 > z'_2$ and such an 
event would survive the first cut. To eliminate these remaining elastic events, we 
consider the \emph{worst-case} scenario when $\theta_{\mathrm{meas}}$ is too large; 
that is, where there are measurement errors of $+\Delta z$ at $R_0$ and 
$-\Delta z$ at $R_1$. In this situation, our ``worst-case'' value of 
$\theta_{\mathrm{true}}$ is computed using $z_0 - \Delta z$ and $z_1 + \Delta z$, 
instead of $z_0$ and $z_1$. We then find the outgoing energy for an elastic 
collision with this scattering angle and repeat the simulation 
to find the new expected position $z''_2$. After cutting all events with 
$z_2 < z''_2$, we are guaranteed that no elastic events have survived. 
For $70^{\circ} < \theta_{\mathrm{true}} < 110^{\circ}$, $\Ebeam = 40~\MeV$, and 
$\Mchi = 2~\MeV$, a simple Monte Carlo simulation shows that the majority of 
dark matter production events survive the cuts.

\subsubsection{Rates and Sensitivity}
\par
The rate of elastic events would scale up from what was
discussed in Section~\ref{S:designI} due to the much larger
solid angle covered.  Also, one would presumably employ a 
more intense beam and/or thicker target.  We imagine that
on the order of $10^{14}$ elastic scattering events would
be recorded, giving a sensitivity on the order of
$10^{-8}~\pb$.  More importantly, a large sample of 
signal events could be detected, allowing a real measurement
of the cross-section and of the mass of the $U$-boson.

\subsubsection{Threshold Behavior}
\par
Fig.~\ref{F:dsigma}~(BOTTOM) shows the excitation curves for
our signal process.  Our default beam energy, $\Ebeam = 40~\MeV$,
is close to the maximum cross-section when $\MU = 10~\MeV$.
With the larger event samples expected in this more sophisticated
experimental design, one could measure the cross-section
as a function of~$\Ebeam$, and trace out the threshold behavior.
Fig.~\ref{F:dsigma} indicates that even just three modest measurements at
$\Ebeam = 40~\MeV$, $20~\MeV$ and $80~\MeV$ would allow one to
distinguish clearly between the three mass values $\MU = 10~\MeV$,
$20~\MeV$ and $5~\MeV$.
Furthermore, experimental design~II would allow a measururement of the angular
variation of the signal process.  As shown in Fig.~\ref{F:dsigma2}~(TOP),
a clear distinction with respect to elastic scattering could be made.

\subsubsection{Other Signatures}
\par
\begin{figure}
\begin{center}
\includegraphics[width=0.45\textwidth]{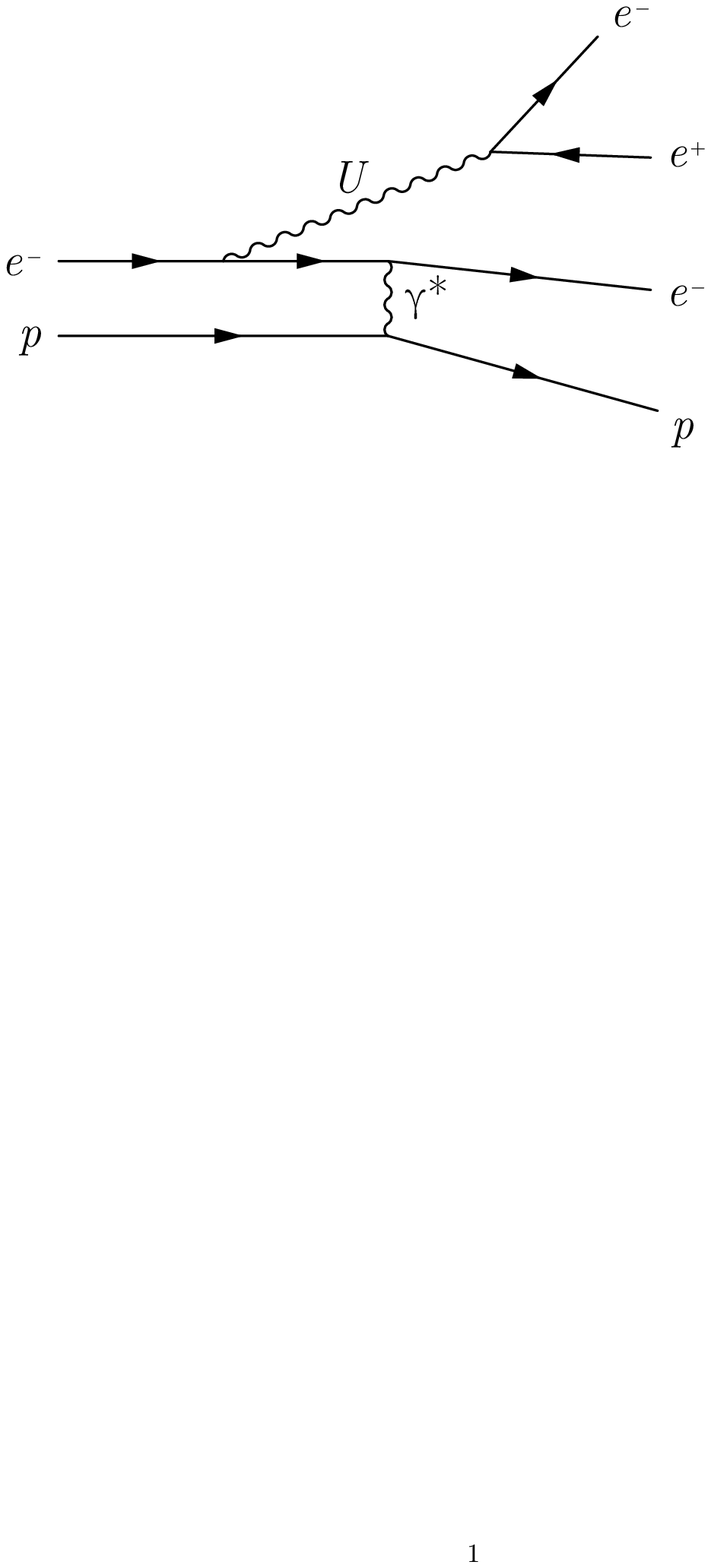}
\caption[.]{\label{F:pairprod}
\em Feynman diagram for $\epem$ pair production via a $U$ boson}
\end{center}
\end{figure}

Returning to the light dark matter model, we recall that the intermediate
particles ($U$-boson and/or $F^\pm$-fermions) couple to electrons, and
hence the process
$$
   e^- p \rightarrow  e^- p \, \epem
$$
must occur (see Fig.~\ref{F:pairprod}).  The invariant mass of the
extra $\epem$ pair would peak at the $U$-boson mass. If the electron
coupling $f_e$  is not too low, then one might obtain a sample of such
events, which would be easy to distinguish from any other process.
The observation and  measurement of all final-state
particles would allow an incisive study of this model.  
The prediction for the rate of this process, however, is more model 
dependent, since it depends on~$f_e^2$ rather than $\Cchi\fe$, and
is less constrained by the dark-matter abundance, Eq.~(\ref{Eq:DMabundance}).

%---------------------------
%---------------------------

%%%%%%%%%%%%%%%%%%%%%%%%%%%%%%%%%%%%%%%%%%%%%%%%%%%%%%%%%%%%%%%%%%%%%%
\subsection{Comparison to $\epem$-Scattering}
\par
A classic study of invisible final states has been carried out
at $\epem$ colliders.  The idea is to tag the event by the
presence of a wide-angle photon radiated from the incoming 
electron or positron.  Recently, Borodatchenkova~\etal\ proposed
using this process to detect light dark matter~\cite{epem}.
Also, Zhu discussed the possibility of making an observation
at BES~III~\cite{Zhu}, and claims a sensitivity down to
$\fe \sim 10^{-5}$.
\par
Our proposal benefits in a number of ways.  First, the luminosity
is much higher due to the use of a fixed target.  Second, the 
kinematics are particularly advantageous, since at tree-level
there are no backgrounds and elastic scattering is highly 
constrained.  Finally, the recoil mass can be used to measure
the $U$-boson mass, in principle.

%%%%%%%%%%%%%%%%%%%%%%%%%%%%%%%%%%%%%%%%%%%%%%%%%%%%%%%%%%%%%%%%%%%%%%
\section{Summary and Conclusions}
\par
The light dark matter model proposed by Boehm and Fayet can explain
both the relic density of dark matter and the 511~keV gamma-ray line
coming from the center of the galaxy.  It survives many constraints
derived from precision measurements in particle physics, and also
several astrophysical observations.  The key ingredients of this model
include a stable light neutral scalar, $\chi$, and a light neutral vector
boson $U$ with a large coupling to $\chi$ and a small coupling to
electrons.
\par
We propose a simple experiment to produce the $U$-bosons in
low-energy electron-proton scattering.  Basically, the $U$-boson 
would be radiated from the incoming electron, resulting an invisible final 
state which can be identified through dramatic kinematical differences with respect to
elastic scattering.  The high luminosity and easy control of the kinematics
should allow the experiment to reach a sensitivity which would confirm or definitively
rule out this model.  Two conceptual designs are described, which will be
modeled and studied in greater detail in the future.

%%%%%%%%%%%%%%%%%%%%%%%%%%%%%%%%%%%%%%%%%%%%%%%%%%%%%%%%%%%%%%%%%%%%%%
\section{Acknowledgments}
\par
We wish to thank C\'eline Boehm and David Buchholz for helpful discussions
in the initial stages of this study, and Pierre Fayet, John Beacom and
David Jaffe for very useful comments after the first version of this
paper appeared.
S.H.\ thanks Edward Boos, Rikkert Frederix and Fabio Maltoni for technical
assistance. The work of S.H.\ was partially supported by CICYT 
(grant FPA2006--02315).

%======================================================================

\end{document}